\AddToHook{begindocument/before}{\usepackage{hyperref}}

\documentclass{aa}
\usepackage[varg]{txfonts}

\date{May 2025}

\begin{document}

\title{A New Method of Deriving Doppler Velocities for Solar Orbiter SPICE}
\titlerunning{SPICE Doppler Velocities}
\author{
    J.~E.~Plowman\inst{\ref{inst1}} \and 
    D. M. Hassler\inst{\ref{inst1}} \and 
    M. E. Molnar\inst{\ref{inst1}} \and
    A. K. Shrivastav\inst{\ref{inst1}} \and
    T.~Varesano \inst{\ref{inst1}, \ref{inst:CUBoulder}} \and
    F.~Auch\`{e}re\inst{\ref{inst2}} \and
    A. Fludra\inst{\ref{i:ral}} \and
    T.~A.~Kucera \inst{\ref{i:gsfc}}\and
    T. J. Wang\inst{\ref{i:cua},\ref{i:gsfc}} \and
    Y. Zhu \inst{\ref{inst:ethz},\ref{inst:pmodwrc}}
    }

\authorrunning{J.E. Plowman Et al.}
\institute{
    Southwest Research Institute, 1301 Walnut St, Suite 400, Boulder, CO 80302, USA \\ \email{jplowman@boulder.swri.edu}\label{inst1}
    \and
    UKRI STFC, RAL Space, Didcot, OX11 0QX, UK \label{i:ral}
    \and
	Université Paris-Saclay, CNRS, Institut d'Astrophysique Spatiale, Bâtiment 121, 91405, Orsay, France\label{inst2}
    \and
    Department of Physics, Catholic University of America, 620 Michigan Avenue, Washington, DC 20064, USA\label{i:cua}
    \and
    Heliophysics Division Goddard Space Flight Center, Greenbelt MD 20771, USA \label{i:gsfc}
    \and 
    ETH-Z\"urich, Wolfgang-Pauli-Str. 27, 8093 Z\"urich, Switzerland \label{inst:ethz} 
    \and
    Physikalisch-Meteorologische Observatorium Davos/World Radiation Center (PMOD/WRC), Dorfstrasse 33, 7260 Davos Dorf, Switzerland \label{inst:pmodwrc}
    \and 
    Aerospace Engineering Sciences, University of Colorado, 3775 Discovery Drive, Boulder, CO, USA. \label{inst:CUBoulder}
}
\date{Received <date> /
       Accepted <date>}
       
\abstract{This paper presents a follow-up to previous work on correcting PSF-induced Doppler artifacts in observations by the SPICE spectrograph on Solar Orbiter. In a previous paper, we demonstrated correction of these artifacts in the $y-\lambda$ plane with PSF Regularization, treating the forward problem with a method based on large sparse matrix inversion. It has since been found that similar apparent artifacts are also present in the $x-\lambda$ direction, i.e., across adjacent slit positions. This is difficult (although not impossible) to correct with the previous matrix inversion method due to the time variation between slit positions. We have therefore devised a new method which addresses both $x-\lambda$ and $y-\lambda$ artifacts simultaneously by applying wavelength dependent shifts at each $x-y$ plane of the spectral cube. This paper demonstrates the SPICE data issue, describes the new method, and shows a comparison with the previous one. We explore the time variation of the correction parameters for the SPICE data and show a clear orbit dependence. The results of the method are significantly higher quality derived Doppler signals, which we estimate at less than $\sim$ 5\,km/s uncertainty for brighter lines in the absence of other systematics. Furthermore, we show the new SPICE polar observation results as a demonstration. The correction codes are written in Python, publicly available on GitHub, and can be directly applied to SPICE level 2 datasets.
}


\maketitle
\section{Introduction}
\label{sec:Intro}

    This paper is a follow-up on a previous publication on correction of point spread function (PSF) artifacts in Solar Orbiter SPICE \citep[][from now on Paper I]{PlowmanEtal_SPICEPSFI}, which contains a more detailed review of the problem; nevertheless we will begin with an abbreviated review and update in this section.

    
    The Spectral Imaging of the Coronal Environment \citep[SPICE,][]{SPICEInstrument_AA2020, FludraEtal_SPICE2021} is an EUV slit scanning spectrograph onboard Solar Orbiter~\citep{2020A&A...642A...1M,SolarOrbiter_AA2021}. It is designed to distinguish between solar wind source models by inferring the elemental composition of solar wind source regions on the surface of the Sun, while also detecting outflow velocities by means of the Doppler shifts of the atomic lines being observed~\citep{2020A&A...642A...3Z}. SPICE was designed in synergy with the in-situ Solar Wind Analyser (SWA) suite of instruments on Solar Orbiter, in order for the instrument's combined observations to trace the evolution of solar wind streams from the surface of the Sun to the Solar Orbiter spacecraft~\citep{2020A&A...642A..16O}.


    SPICE observes the solar spectrum in two wavelength windows, from 704 to 790\,\AA\ and from 973 to 1049\,\AA\ ~\citep{SPICEInstrument_AA2020}. The spectral lines at these wavelengths are sensitive to solar plasma temperatures ranging from 10\,kK to 1.0\,MK, with the addition of two hotter Fe XVIII and Fe XX lines that can be seen only in flares~\citep{SPICE_FEXX_lines}. These windows were chosen to maximize the selection of complementary (similar temperature sensitivity yet differing atomic ionization susceptibility) line pairs for measuring atomic abundance in the solar wind source regions \citep{VaresanoEtal_SPICEMosaics_AA2024}.


    Although abundance measurements are the primary driver for SPICE, Doppler velocity measurements (i.e., fitting the profiles of the spectral lines in wavelength and thereby identifying the mean velocity of the emitting regions via the Doppler effect) are crucial for the maximal scientific output of the mission. It is a key piece of information in identifying solar wind source regions \citep{HasslerEtal1999, 2008Harra_AR_outflows}, and such Doppler velocity measurements provide initial conditions for understanding the driving mechanisms and evolution of the solar wind \citep{2009Baker}. Additionally, accurately determining the properties of overlapping spectral lines (whether blended, or merely near each other in wavelength) is dependent on the spectral veracity of the data. 

    A further advantage of SPICE observations is that, in conjunction with Earth-perspective spectral observatories such as the Hinode EUV Imaging Spectrograph \citep[EIS,][]{HinodeEIS} and the Interface Region Imaging Spectrograph \citep[IRIS,][]{DePontieu_IRIS_SoPh2014}, they can provide multi-viewpoint spectral and Doppler shift observations of solar features of interest. This is an entirely new capability offered by Solar Orbiter, and leveraging this capability requires well calibrated Doppler shift observations.

\begin{figure*}[!htbp]
        \begin{center}
            \includegraphics[width=\textwidth]{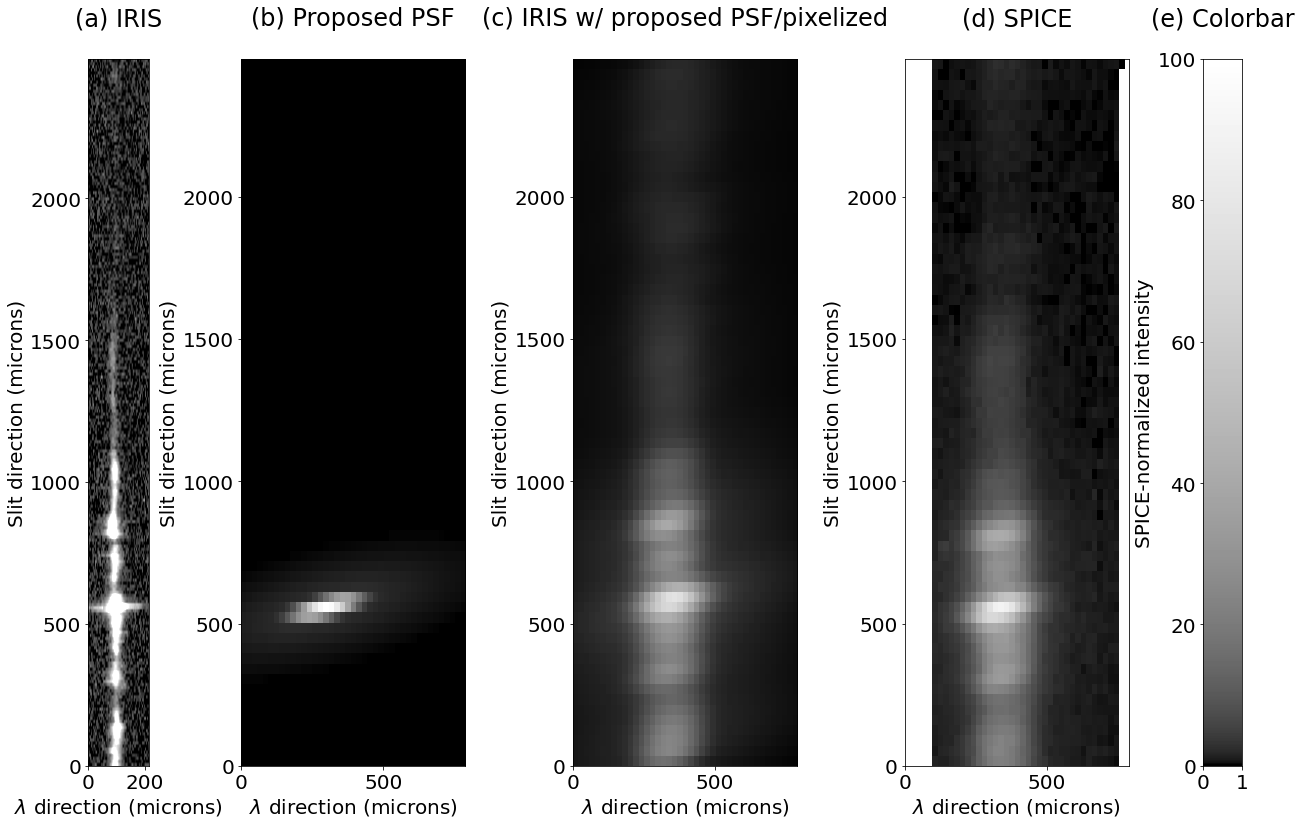}
        \end{center}
        \caption{Illustration of the SPICE PSF issue in the $y-\lambda$ plane. The PSF is tilted in the $y-\lambda$ plane as shown in panel (b), which leads to leakage of signal from the bright features to adjacent pixels, not strictly at the same $y$ or $\lambda$ values, as shown in panels (c) and (d). Measurements of Doppler velocity of the data in panels (c) and (d) would lead to artificial Doppler shifts,  as shown in Figure~\ref{fig:initial_psf_issue_illustration_Doppler}.}
        \label{fig:initial_psf_issue_illustration}
    \end{figure*}
\begin{figure*}[!htbp]
        \begin{center}
            \includegraphics[width=\textwidth]{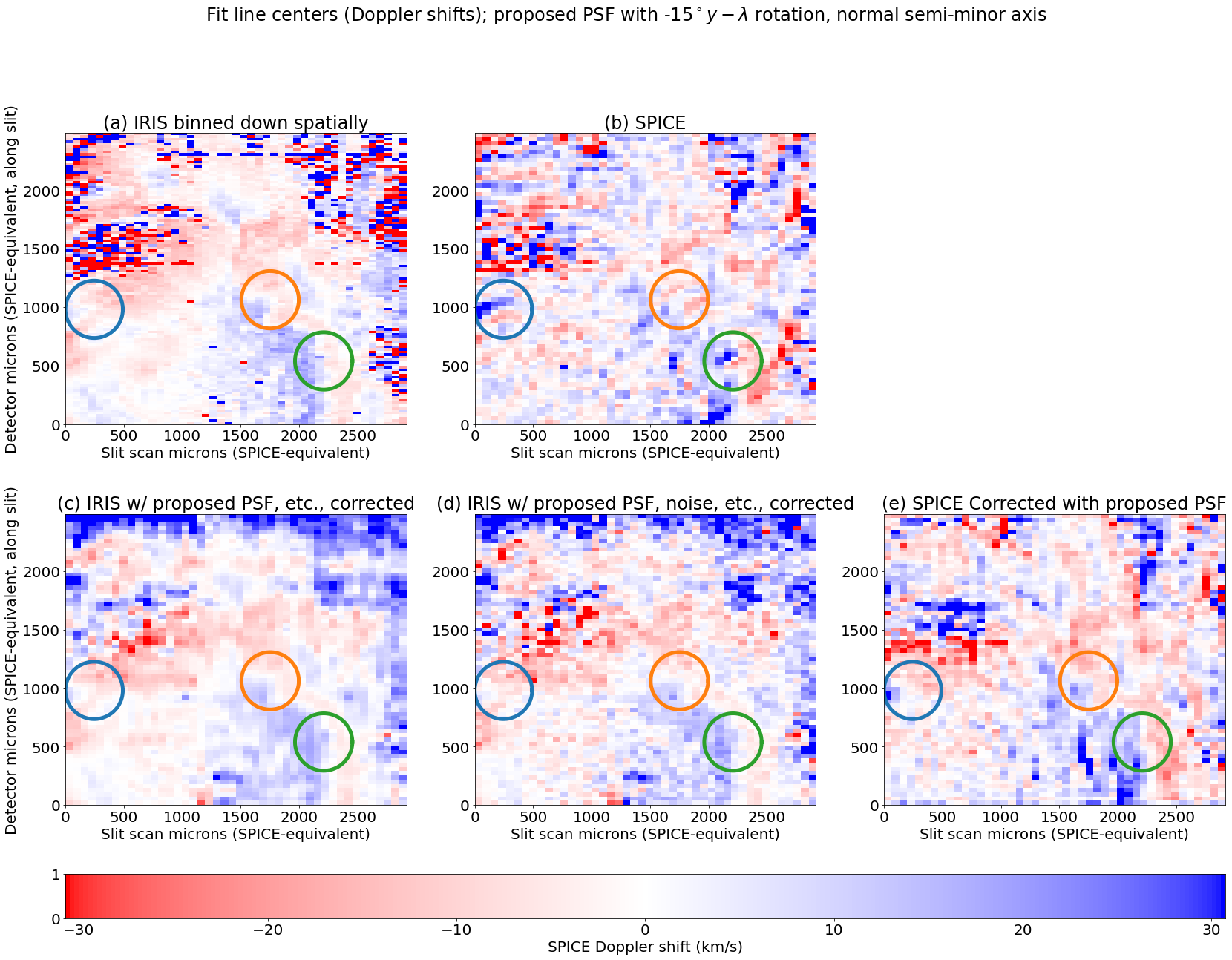}
        \end{center}
        \caption{Illustration of the Doppler velocity lobes caused by the SPICE PSF artifacts from a co-observing campaign between IRIS and SPICE. The SPICE data show a characteristic redshift and blueshift pattern (examples circled) around bright features in intensity -- compare the top left panel (IRIS) with top right panel (SPICE) observing the same location. This pattern similarly appears in IRIS (not shown) when the PSF shown in Figure \ref{fig:initial_psf_issue_illustration} is applied to IRIS (not shown) and (largely) disappears when the PSF regularization method described in \citet{PlowmanEtal_SPICEPSFI} is applied to it. This is strong evidence that the Doppler velocity lobes are caused by the PSF tilt out of $\lambda$ and into $y$.}
        \label{fig:initial_psf_issue_illustration_Doppler}
    \end{figure*}

    Early in the Orbiter mission, it was found that SPICE Doppler velocity maps tend to show a pattern of elongated, overly large Doppler shifts around bright features \citep{FludraEtal_SPICE2021}. An inspection of the spectra (see Figures \ref{fig:initial_psf_issue_illustration} and \ref{fig:initial_psf_issue_illustration_Doppler}) showed that an apparent tilted PSF feature in the $y-\lambda$ plane is responsible. This causes light from bright features to leak into adjacent spatial pixels at {\em offset} wavelengths. 
    To correct this issue, we developed a sparse matrix correction method, which works by effectively deconvolving (although by some definitions a deconvolution involves a spatially invariant kernel, whereas the sparse matrix method does not have this limitation) the PSF \citep{PlowmanEtal_SPICEPSFI}. We then reapplied a nominal PSF, making it a PSF regularization approach (i.e., we first remove the aberrant PSF then we reapply the nominal PSF so that the resulting data is free of noise and sharpening artifacts).
    The results of this procedure, as well as example PSF artifact lobes, are shown in Figure~\ref{fig:initial_psf_issue_illustration}.

\begin{figure*}[!htbp]
        \begin{center}
            \includegraphics[width=\textwidth]{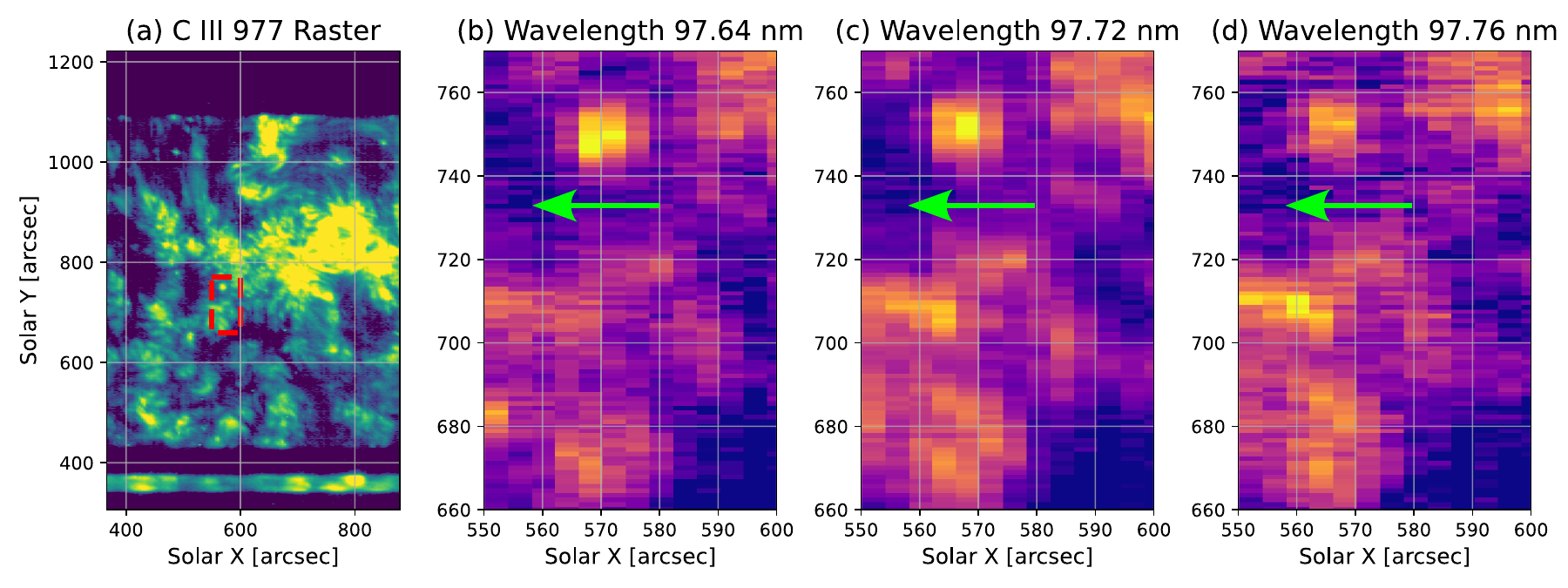}
        \end{center}
        \caption{Illustration of the $x-\lambda$ and $y-\lambda$ shifts in a slice of a SPICE spectral cube, with the intensity raster shown in the C III 977\,\AA\ line. Panels (b), (c), and (d) show zoomed-in slices (corresponding to the red square in panel (a)) where the $x-y$ shift is apparent as the slight translation of the same features, when scanning through consecutive wavelengths. Please see the attached video for an illustration of this effect. }
        \label{fig:xlambda_psf_issue}
    \end{figure*}

    It was more recently discovered (P. Young, 2023, private communication) that an apparently similar issue aliases bright features into $x$ (i.e., the slit scan direction, compared to the $y$-direction corresponding to the one along the slit) at differing $\lambda$ locations. This was originally discovered by noticing that when a spectral cube sequence is sliced through wavelength, an apparent image-wide shift is observed, as shown in Figure \ref{fig:xlambda_psf_issue}. This is an equivalent way of visualizing such issues as the $y-\lambda$ image shown above, and a $x-\lambda$ image can show a similar tilt to that in the $y-\lambda$ image. It should be noted that $x$-direction is built up by rastering the perpendicular to the slit direction by stepping from the primary mirror tilt
    while only the $y-\lambda$ plane is instantaneously imaged directly on the detector. 
    
    The presence of this kind of effect in $x-\lambda$ images was not initially expected since it was thought that the presence of the slit would confine such spatial-spectral crosstalk to the slit plane. Additionally, some observations initially considered did not manifest a detectable $x-\lambda$ shift or significant Doppler shift artifacts, which contributed to it being overlooked (we will return to the temporal variation of these effects later in the paper). Nevertheless, since the behavior of the $x-\lambda$ shift appears otherwise equivalent to that in $y-\lambda$ direction we treat it -- ``to the same natural effects we must, as far as possible, assign the same causes'' \citep{1687pnpm.book.....N} -- as also being caused by the general PSF shape. Referring to the optical layout of SPICE \citep[Figure 3 of][]{SPICEInstrument_AA2020}, we note that the diffraction grating is downstream from the slit, so the slit ought to isolate the observations from grating-induced $x-\lambda$ crosstalk. The mirror enclosure, containing the scanning mechanism, however, is upstream of the slit; so this would seem to implicate the mirror as the source of this effect. It is not exactly clear what the physical mechanism is for this optical aberration at this point,  although the mirror is subject to considerable thermal stress at perihelia. In any case, it is not the purpose of this paper to fully explain the effect but rather to show how it can be corrected, which we turn to next.

    First, we consider the use of the previously developed method in Paper I. It was specifically built to permit multi-dimensional PSF correction, and it is in fact capable of doing so. However, there are two significant issues that make it quite difficult to apply the full 3D PSF correction using this method:
    \begin{enumerate}
    
        \item It has proven quite challenging to find a sufficiently accurate PSF model applicable for different times during the mission. In principle, given perfect knowledge of the instrument, it could be calculated, but we are missing detailed information about the state of the instrument and understanding of the origin of this optical aberration to be able to predict the instantaneous 3D instrumental PSF. Instead we are left with an unknown time-varying 3D PSF that has a vastly larger parameter space to search than a 2D one. Additionally, the correction itself is much more computationally demanding in 3D than in 2D, so the compute time to search the 3D PSF space is orders of magnitude more than for the 2D PSF space.
        
        \item Even if ignoring point 1, the $x$ direction is obtained by rastering in time. This means that each $x$ slice is at a different time than the others, and since the solar surface changes on temporal scales similar to the scanning timesteps, the different raster steps are observing different solar sources. The forward matrix inversion or deconvolution-based schemes assume that the instrument is observing the same source throughout the whole rastering process. Anywhere that assumption is broken, the corrected image will have large spikes when the algorithm attempts to describe a temporal jump with an incorrect PSF. This could be addressed by explicitly adding a 4th dimension, time, to the inversion. However, the data lacks sufficient information to fully disambiguate the temporal and spatial $x$ directions, while again adding significant computational overhead. 
    \end{enumerate}

    Instead, we have found that a simpler method, based on corrective image shifts (or skews, depending on which axis of the spectral cube is being viewed), can jointly address PSF artifacts in both the $x-\lambda$ and $y-\lambda$ directions. Although this approach lacks the rigor and completeness of the previous reconstruction method, and lacks the improvement of spatial and spectral resolution that the previous method offers, it appears to work as well or better than the previous method for reducing PSF artifacts and it is far less computationally demanding, without knowing the exact instantaneous shape of the PSF \textit{a priori}. We now turn to describing and demonstrating this method.

\section{Description of new PSF Correction Method}

\begin{figure}[htp!]
    \centering
    \includegraphics[width=\linewidth]{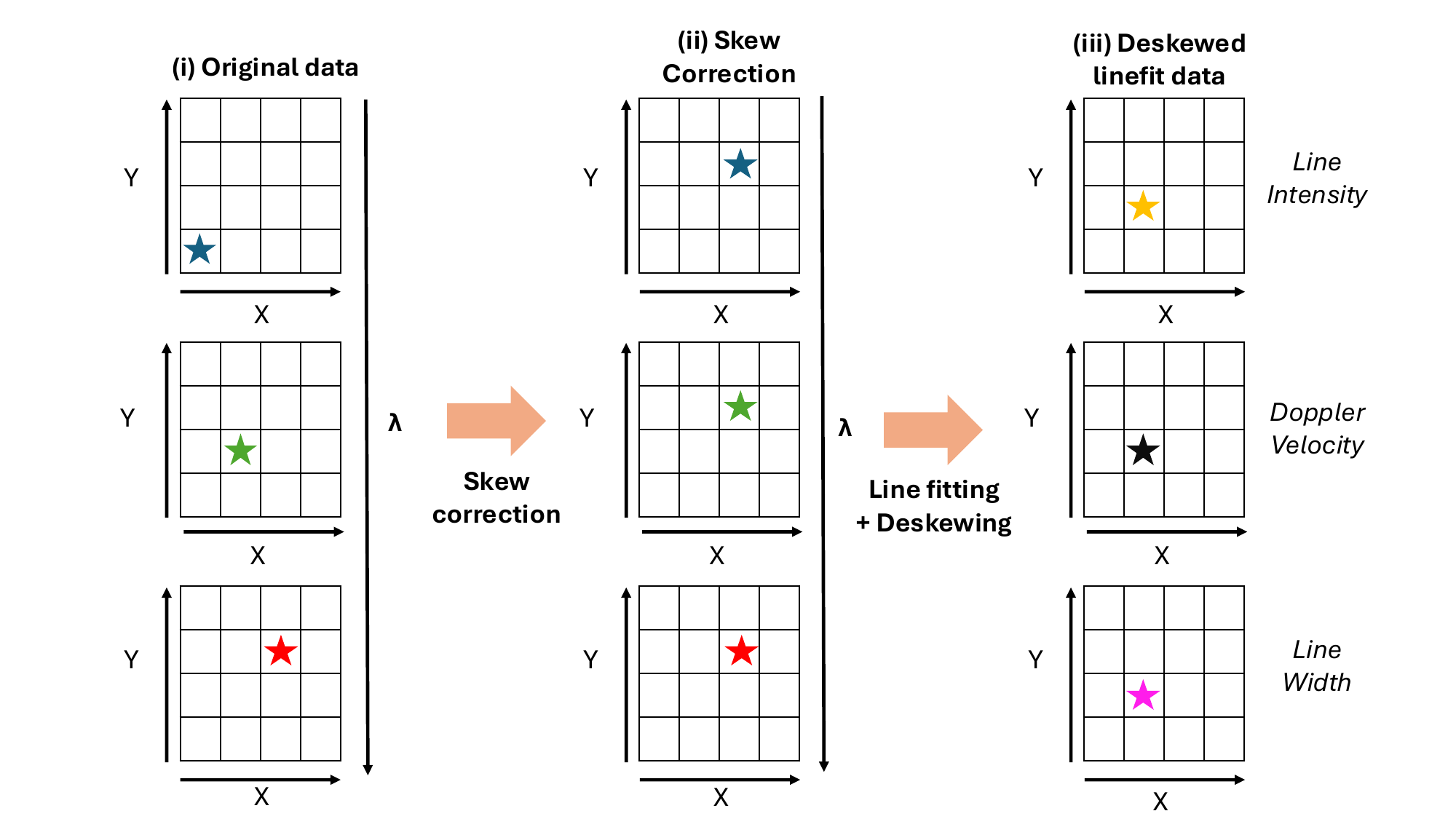}
    \caption{Schematic of the algorithm. The original data in the left panel shows a wavelength-dependent shift. After the application of the skew correction in the middle panel, the wavelength shift/skew is removed, but a leftover shift is present. The bottom row of this figure corresponds to $\lambda_0$ since the skew doesn’t shift the star at that wavelength pixel. A final deskewing or `de-warping' is needed because the correction introduces small spatial shifts to the profile of each spectral line that are generally different at each pixel. These deskewed line profile parameters are the final processed data (right panel). The third (rightmost) column of the figure has only one panel to reflect that the deskew is performed on spectral line fits which have line parameters instead of a $\lambda$ axis.}\label{fig:Algorithm_figure}
\end{figure}

    Since the PSF tilt appears as a shift when scanning through wavelength (see panel i) of Figure~\ref{fig:Algorithm_figure}), we apply a compensating wavelength dependent spatial shift to each $x-y$ plane (constant wavelength plane) of the spectral cube. For example, if each pixel of an $x-y$ plane in the input cube at wavelength $\lambda_k$ is at coordinates $(x_i, y_j)$, we reinterpolate the output cube to new pixel coordinates $(x_i', y_j')$ given by:
    \begin{equation}\label{eq:shift_equation}
        x_i', y_j' = x_i+f_x(\lambda_k), y_i+f_y(\lambda_k)
    \end{equation}
    After testing, we found that a linear function for the shifts per unit wavelength ($f_x(\lambda_k), f_y(\lambda_k)$) from wavelengths $\lambda_k$ suffices:
    \begin{equation}\label{shift_deltas}
        f_x(\lambda_k), f_y(\lambda_k) = \Delta x (\lambda_k-\lambda_0), \Delta y (\lambda_k-\lambda_0),
    \end{equation}
    
    where $\lambda_0$ is a reference wavelength (usually the center of the spectral window or reference wavelength for the line in question) and $\Delta x$, $\Delta y$ are the constant coefficients of the shifts. With this approach in place, counteracting the PSF effects on the Doppler shifts is a simple matter of reinterpolating the image to the new coordinates, producing the results in panel ii) of Figure~\ref{fig:Algorithm_figure}. We use the \texttt{scipy.interpolate.RegularGridInterpolator}  method which is a part of the \texttt{scipy} scientific Python programming package \citep{scipy} for performing the skew correction. The default settings for the interpolation interpolation mode are used. 
    
    {
    The wavelength-dependent shifting introduces (typically) subpixel spatial shifts in the image depending on the wavelength of the spectral line compared to the reference wavelength. That is to say, if a spectral line in a particular spatial pixel is centered at wavelength $\Delta \lambda$ compared to the reference wavelength, it will have its image position shifted by $\Delta x \Delta \lambda$ and $\Delta y \Delta \lambda$ in x and y respectively. This can be reversed in the spectral line fits by reinterpolating the fit found at each pixel based on its line center wavelength, as found by the fitting procedure -- i.e., for each spatial pixel, we perform a final spatial shift of $-\Delta x \Delta \lambda$ and $\-\Delta y \Delta \lambda$. This is a non-homogeneous `image-warping' (or `de-warping', really) interpolation step, shown in panel iii) of Figure~\ref{fig:Algorithm_figure}.
    } 
    
    
    This approach has significant advantages over the previous method from Paper I, as described below:
    
    \begin{enumerate}
        \item This correction method can be executed very quickly (a matter of minutes on a single CPU on a modern computer for a single SPICE raster), so the computational overhead of applying the correction is negligible compared to the other components of spectral line analysis (primarily the line fitting of the spectral profiles). 
        \item Unlike the PSF-deconvolution-like method based on large sparse matrix inversions, this method has minimal susceptibility to time variation between raster positions -- only a small temporal smoothing is introduced.
        \item Due to the speed and simplicity of this method  (only two parameters $\Delta x$ and $\Delta y$ need to be optimized), a brute force search for the optimum parameters is practical and achievable with minimal computational cost. We have implemented and carried out such a search which we will describe in subsequent sections.
        
    \end{enumerate}

    However, there are tradeoffs compared to the regularization approach from Paper I, which we think are modest compared to the advantages:
    First, there is no enhancement of spatial resolution, which the PSF regularization from Paper I can offer.
    To the opposite, the multiple layers of interpolation involved in the new method lead to a spatial resolution decrease of no more than about one pixel, roughly a factor of 1.5. Furthermore, the dewarping (last step of the algorithm in Figure~\ref{fig:Algorithm_figure}) has some computational overhead, although still negligible when compared with the 3D PSF deconvolution problem. 
    
\begin{figure*}[!htbp]
        \begin{center}
            \includegraphics[width=\textwidth]{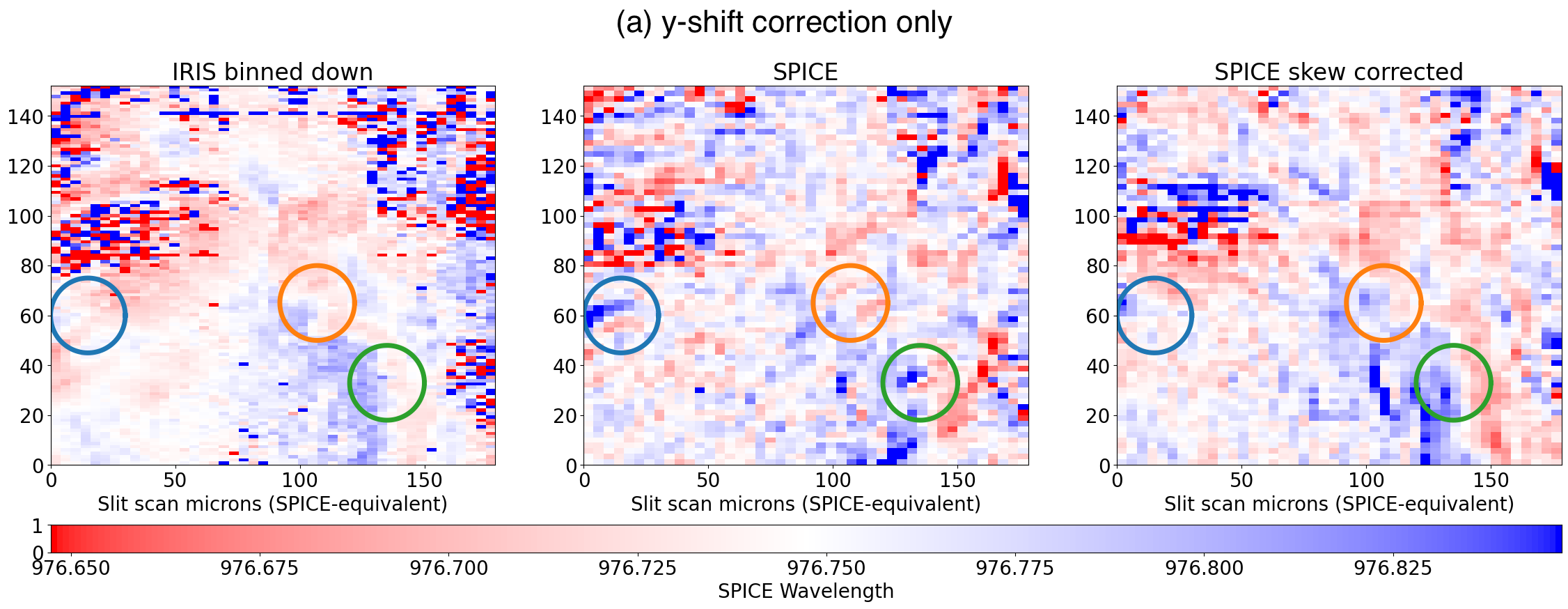}
            \includegraphics[width=\textwidth]{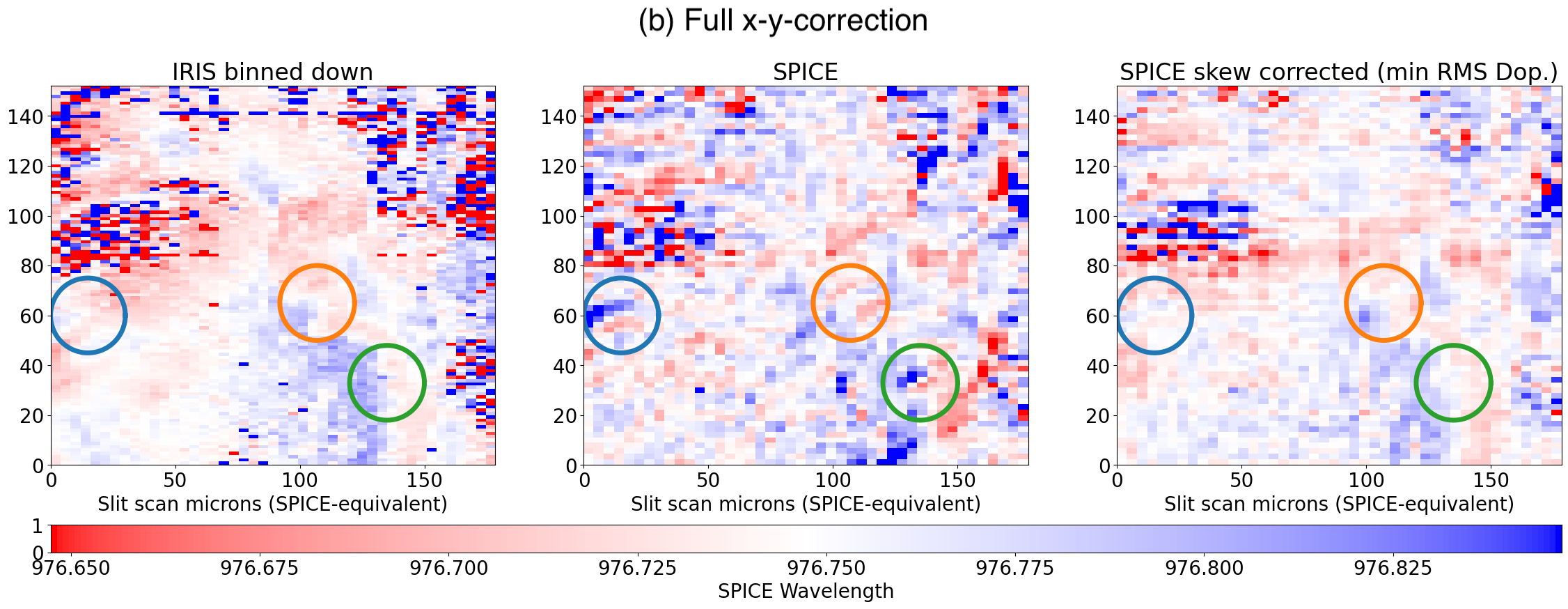}
        \end{center}
        \caption{The new correction method applied to the same data set as in Figure \ref{fig:initial_psf_issue_illustration_Doppler}. The top row (panels (a)) shows the result from a $y$-only wavelength shift applied with the same parameters as those used in the aforementioned figure, while the second row (panels (b)) shows the results with both $x$ and $y$ wavelength-dependent shifts. Notably, including only the $y$-shift shown in panels (a) does not fully remove the Doppler velocity artifacts. Compared to that, including the full $x-y$ shift results in very good agreement between the IRIS and SPICE observations. 
        }\label{fig:new_correction_initial1}
    \end{figure*}

    We have implemented the correction algorithm, including the final dewarping reinterpolation using the \texttt{scipy.interpolate.LinearNDInterpolator} method, with methods from \texttt{numpy} and \texttt{scipy} \citep{scipy}. The results obtained from the application of the new method to the SPICE level-2 dataset are shown in Figure~\ref{fig:new_correction_initial1}. The resolution loss due to the interpolation does not appear to be visually noticeable compared with the previous correction (see Figure~\ref{fig:initial_psf_issue_illustration}).  When equivalent correction parameters between the old and the new correction methods are used (ie., only considering y-$\lambda$ correction), the results look extremely similar, as visually presented in the comparison between Figures~ \ref{fig:initial_psf_issue_illustration_Doppler} and \ref{fig:new_correction_initial1}. However, we found that by applying both x and y shift corrections, the obtained result matches that of IRIS even better. 


\section{Estimation of correction parameters}
\label{sec:Estimation_params}
    To apply the aforementioned method to new SPICE data, the shift parameters 
    $\Delta x$ and $\Delta y$ need to be determined. These parameters can't be readily estimated from the known instrument configuration or calibration data since neither provides sufficient information; potentially, stellar observations may provide an occasional one-point check of PSF aberrations, but these data are only taken occasionally. We presage subsequent parts of the paper (see Section~\ref{sec:Results}) to note that there is significant time variation in this effect and potentially a wavelength-dependent variation as well. Hence, an occasional calibration observation will not suffice, likewise for the hand-tuned set of parameters used in the previous correction paper. Instead, we propose an automated parameter search that produces good results.

    In order to carry out an automated shift parameters search, a figure of merit is required to quantify how well a given parameter set performs. We have found that the standard variation of the Doppler shifts in the line fits (post applying the correction with a proposed set of parameters) appears to produce a high veracity figure of merit. The rationale for this is that the PSF aberration will produce artificial apparent Doppler signals where the Sun exhibits none, but is quite unlikely to remove any existent ones. We use this figure of merit, 
    after subtracting a linear trend in the Doppler shift across both detectors (SPICE appears to show such a trend in its observations, and there is also the linear trend imparted by the solar rotation); this is a planar trend fit over the entire image.
    We only include pixels with estimated Doppler uncertainty (see Appendix~\ref{App:Jacobi_errors} for uncertainty estimation) less than half the estimated Doppler shift, relative to the median wavelength after detrending.

    To find the best correction parameter, we use a combination of brute force and an adaptive grid search. An initial grid of $5\times5$ points is laid out covering the range of shift parameters from $-5$ to $5$ arcseconds per Angstrom. These grid dimensions are chosen to cover the typical range of the parameters. We then check each of the 25 parameters on the initial grid. Following that, we refine the results for this initial grid, starting with a higher resolution grid covering the same range but with $11\times 11$ points. We do not compute the figure of merit for all points on this grid, since we are only interested in the parameter space with the smallest figure of merit. Instead, we interpolate the results for the initial grid onto this new grid, and then we compute the figure of merit for the 20 grid points that have the lowest (interpolated) figure of merit. These will be focused on the lowest point(s) of the merit function found so far. Finally, we repeat this refinement step with a $31\times 31$ point grid, incorporating the points at both earlier grid spacings and computing the figure of merit for the most suitable 20 grid points.
    
\begin{figure*}[!htbp]
        \begin{center}
            \includegraphics[width=\textwidth]{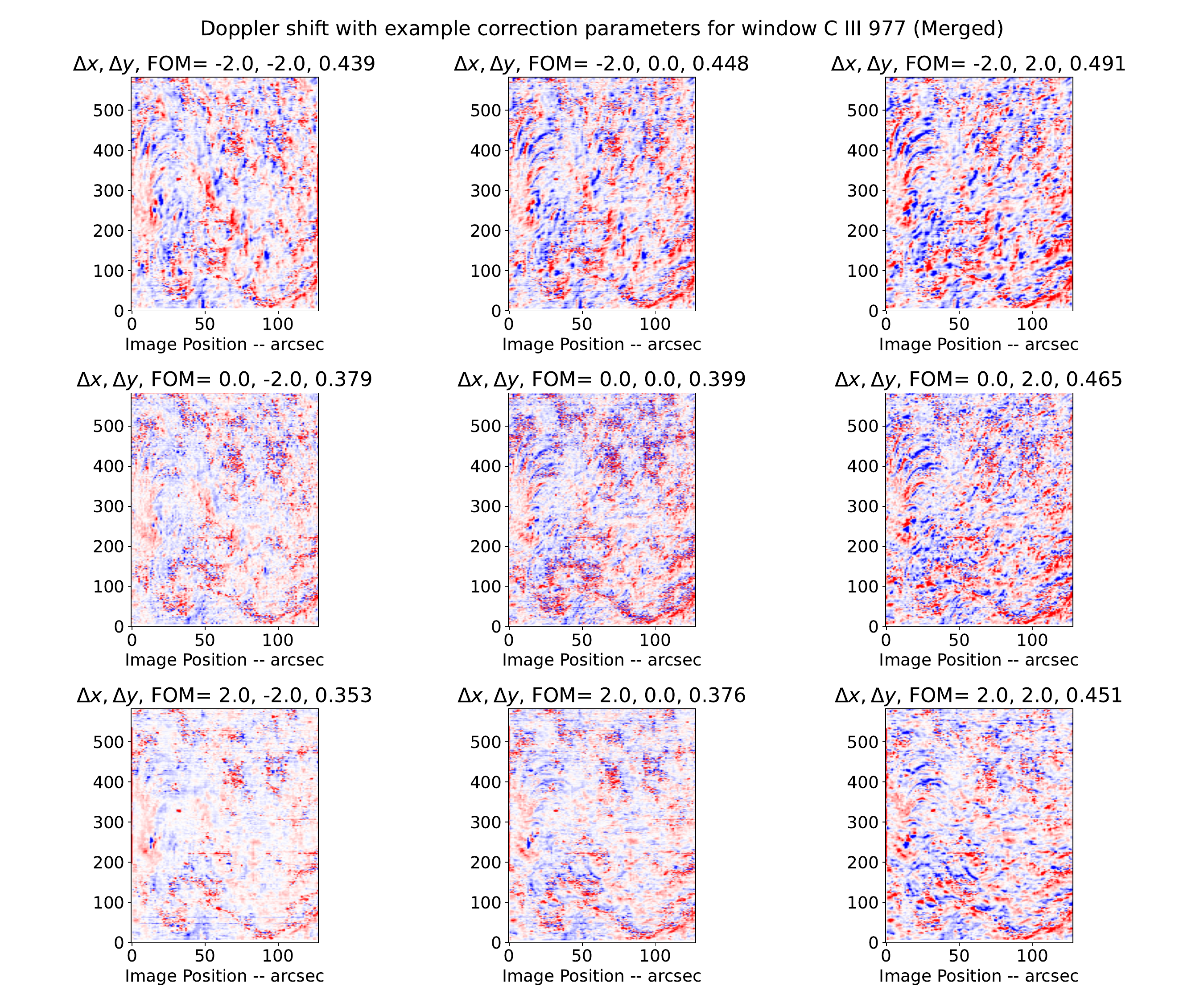}
        \end{center}
        \caption{The detrended Doppler shift signals, on which the figure of merit is based, for the SPICE observations used in the previous PSF work in Paper I and Figure~\ref{fig:new_correction_initial1}. The subpanels show the resulting Doppler shifts from different shift parameters, noted above each subplot. With the highest quality Doppler shift correction around $\Delta x = 2$\,arcsecond/\AA\ and $\Delta y = -2$\,arcsecond/\AA\, this figure is consistent with the automated search determination of $[2,-1.667]$ as shown in Figure~\ref{fig:FOM_surface}.}
        \label{fig:correction_parameter_grid}
    \end{figure*}

    To demonstrate the performance of this search method and the proposed figure of merit, we show in Figure \ref{fig:correction_parameter_grid} the detrended Doppler signals that the figure of merit is based upon, for a grid of $\Delta x$ and $\Delta y$ parameters. The patterns in the Doppler shifts resulting from  different shift parameters (noted as the subplot titles) are clearly apparent, as are the characteristic blue and redshift signatures of the Doppler artifact for some correction parameters, such as those which are in the opposite direction of the required correction. The figure shows the full frame of the same data used in comparison to IRIS in Figure~\ref{fig:initial_psf_issue_illustration_Doppler} and \ref{fig:new_correction_initial1}, which show a subfield in the upper left of Figure~\ref{fig:correction_parameter_grid}. For this dataset, the shift parameters of $\Delta x=2$, $\Delta y=-2$ (arcseconds per \AA\ ) clearly indicate both the least Doppler shifts overall as well as the best correspondence with IRIS by reference to the previous figures. The parameter search figure of merit is the standard deviation of the Doppler shifts, shown in each of these panels. It should be noted that the plots suppress high uncertainty pixels with a soft rolloff rather than the hard cut used to compute the figure of merit.

    \begin{figure*}[!htbp]
        \begin{center}
            \includegraphics[width=\textwidth]{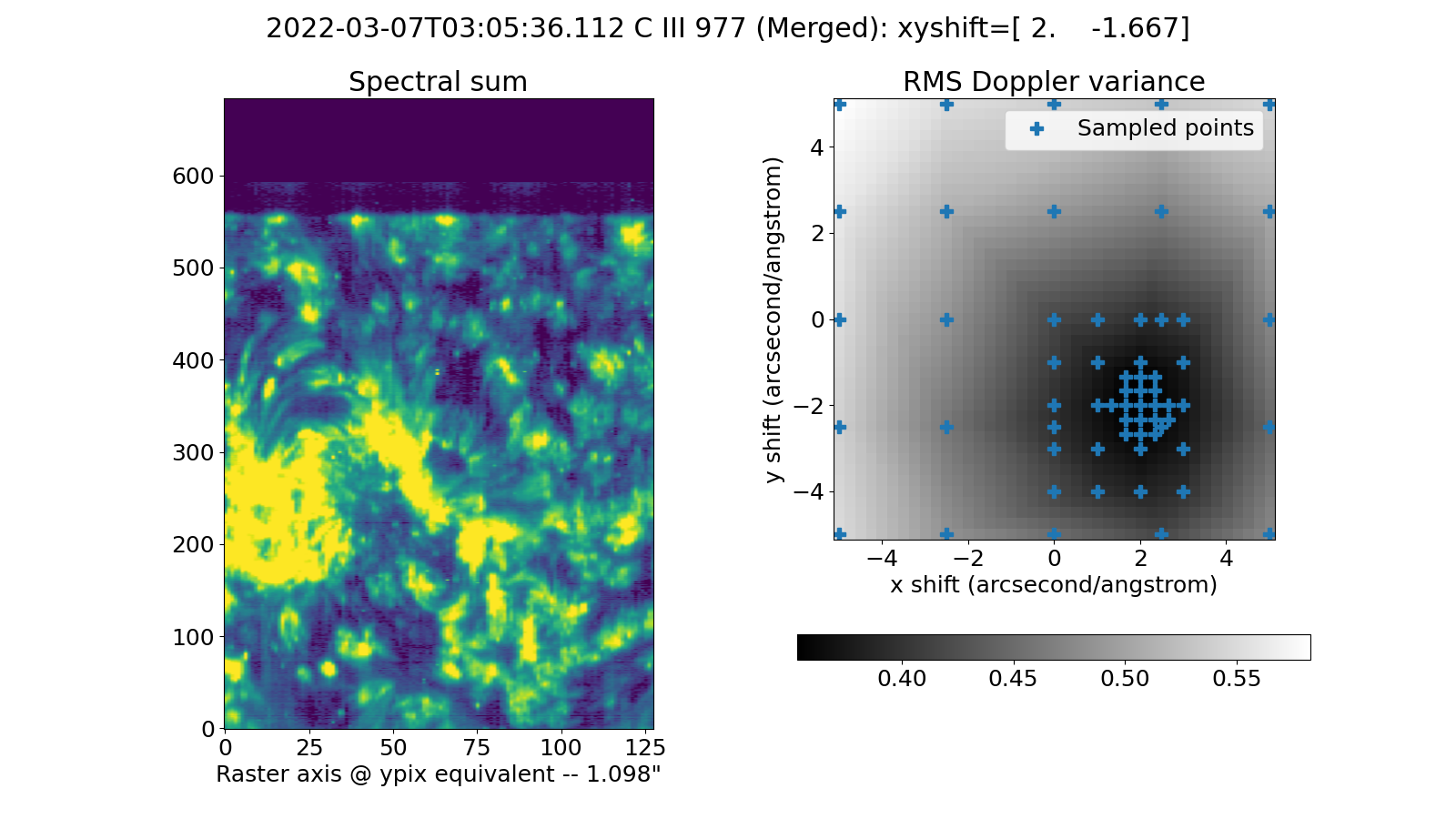}
        \end{center}
        \caption{The figure of  merit surface in $x-\lambda$ and $y-\lambda$ shift parameter for the SPICE observations obtained during the IRIS coordinated observing campaign used for the previous PSF estimation efforts. The right panel shows the figure of merit surface while the left one shows the spectral sum of the window for comparison in the C III 977 Å line (The spectral sum is calculated at each pixel by summing the intensity at all wavelengths in the spectrum).  The blue points in the right panel are where the search procedure checked the figure of merit. The figure of merit shows a clear preference for a $x-\lambda$ as well as a $y-\lambda$ shift, which is confirmed by the better match to IRIS observations.}
        \label{fig:FOM_surface}
    \end{figure*}
    Figure~\ref{fig:FOM_surface} further illustrates the figure of merit surface as functions of $\Delta x$ and $\Delta y$, as mapped out by the automated search algorithm. For the C III 977 Å observation shown in the left panel, used previously throughout the paper, we have computed the figure of merit over the parameter shift space, shown in the right panel. The iterative refinement search algorithm works as anticipated by the pattern of the points being sampled, and it is evident that the merit function surface is fairly smooth. The best parameter pair was found to be $\Delta x=2$, $\Delta y=-1.67$, consistent with the previously shown results in Figure~\ref{fig:correction_parameter_grid}. We note, that these parameters were also used for the comparison with the IRIS data in Figure~\ref{fig:new_correction_initial1}, producing the excellent correspondence between the data from the different instruments.

    As previously mentioned, this result is derived without attempting to specifically tune the correction or the parameter search to recover the IRIS observations. The search algorithm was only optimizing the standard deviation of the detrended Doppler shifts in SPICE, with no reference to the IRIS observations. The IRIS observation comparison was only carried out after the new algorithm has found its preferred candidate parameter pair. Therefore the improved correspondence with the IRIS data, compared to the previous method or the $y-\lambda$ only correction, is itself a validation of the method. 

\begin{figure*}[!htbp]
        \begin{center}
            \includegraphics[width=0.4\textwidth]{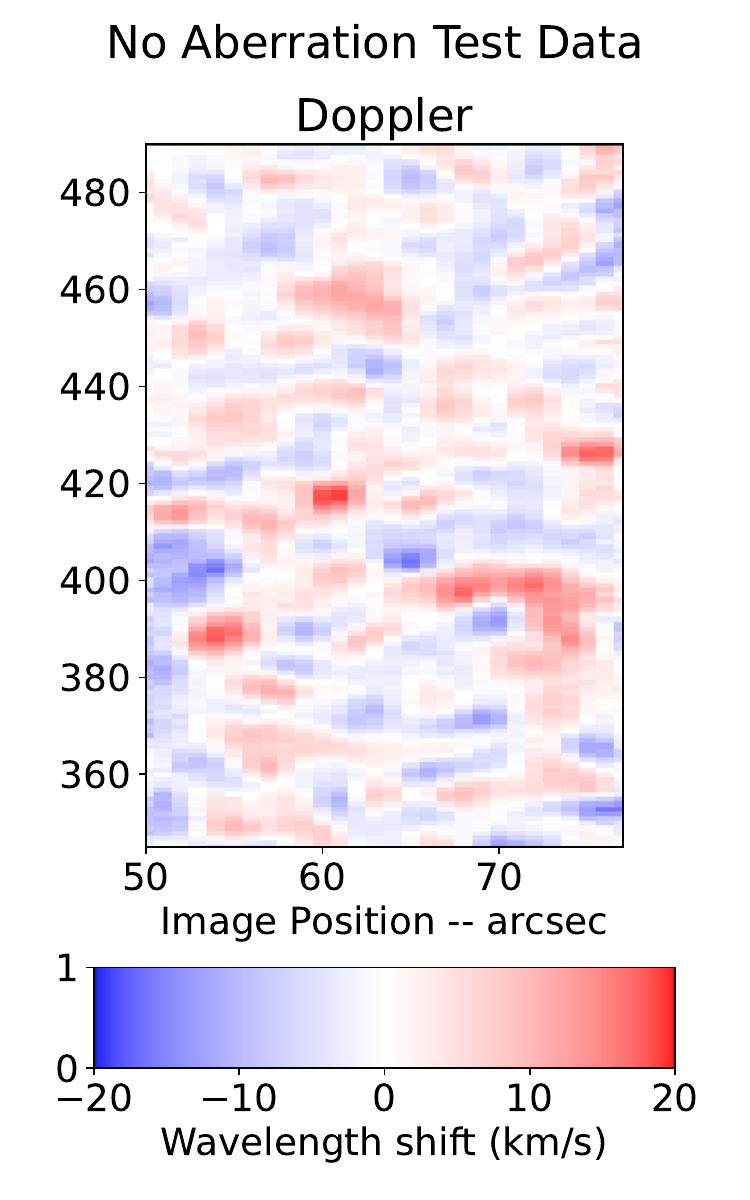}\includegraphics[width=0.4\textwidth]{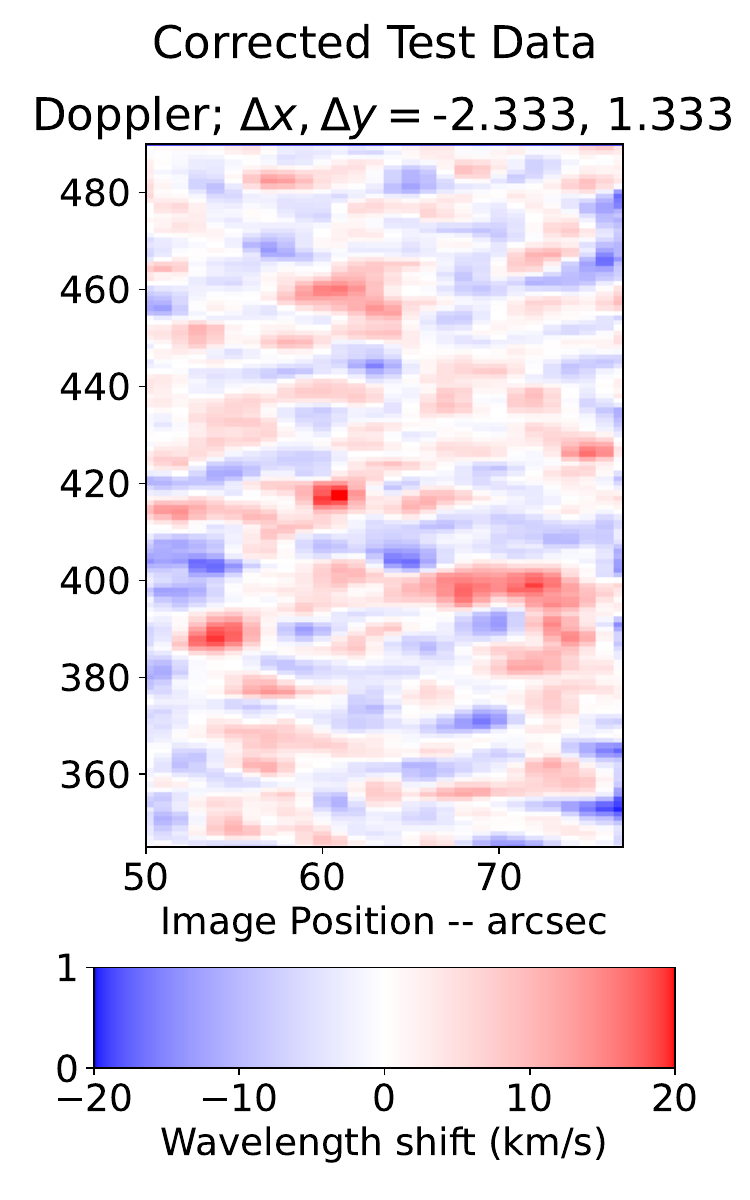}
            \includegraphics[width=0.2666\textwidth]{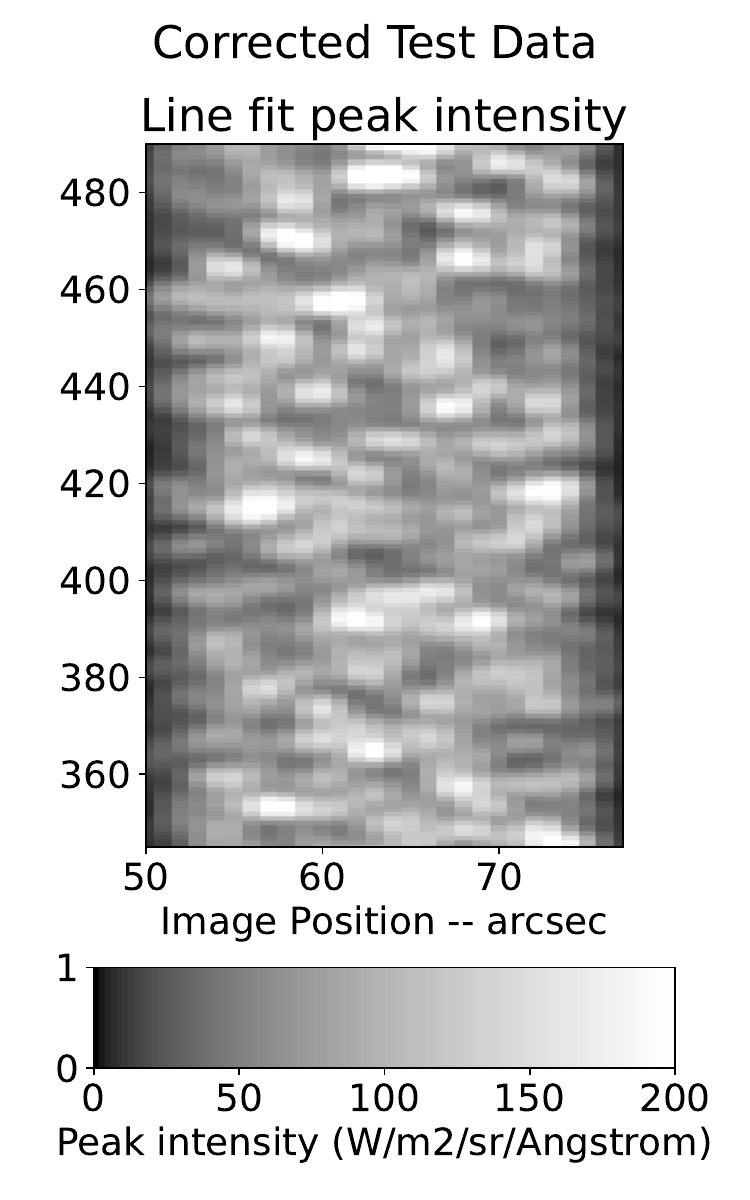}\includegraphics[width=0.2666\textwidth]{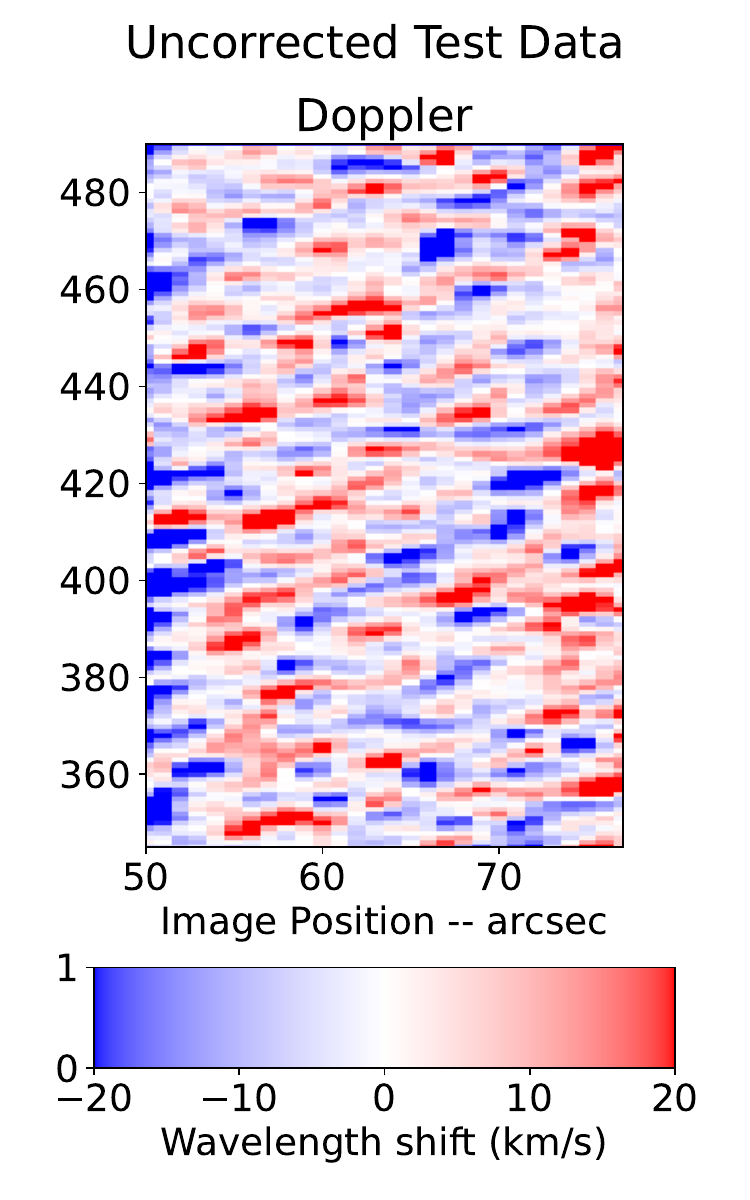}\includegraphics[width=0.2666\textwidth]{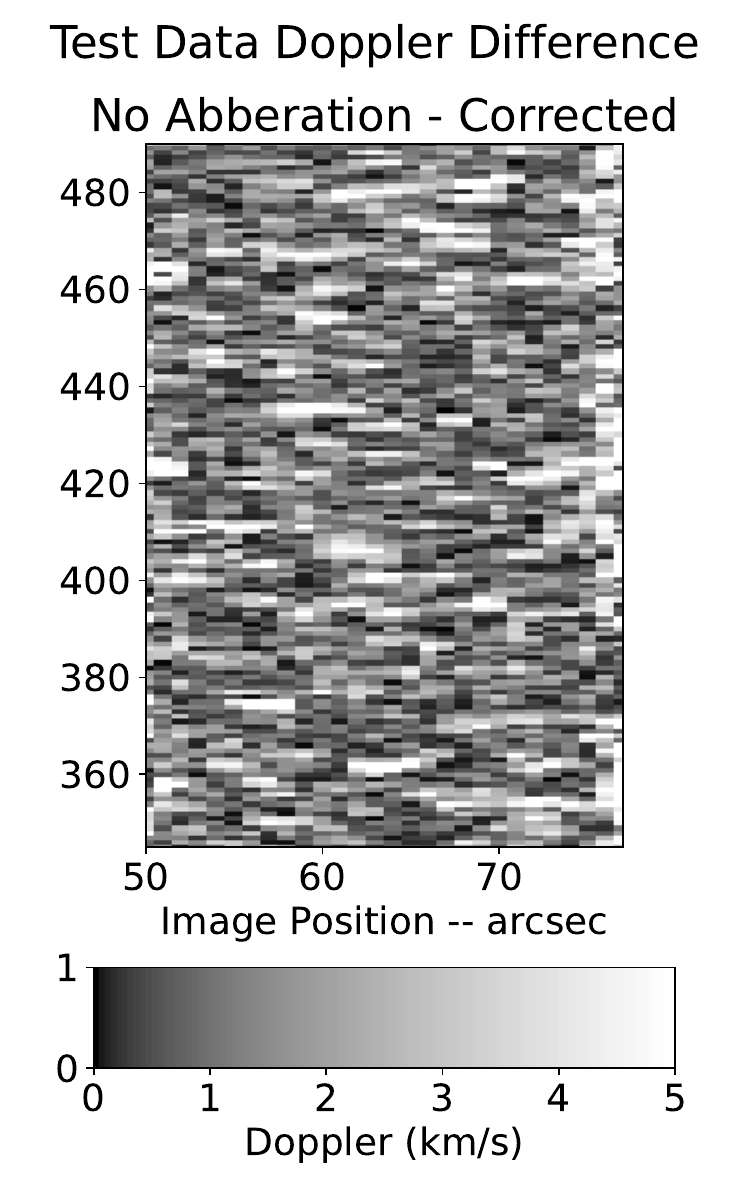}
            
        \end{center}
        \caption{Results of an `end-to-end' test of the new method using random test data. At top left, the Doppler shifts with a nominal (no aberration) PSF is shown, while at top right the Doppler shifts with an aberrant PSF and correction is shown, produced from the same underlying data. The correction recovers the test results well -- at lower right the magnitude of the difference between the two is shown to be under 5 km/s except for very low signal regions adjacent to high signal ones. The uncorrected Doppler shifts (lower center) shows far larger differences, overwhelming the actual signal in this test data set. The line fit intensity is shown at the lower left for reference.}
        \label{fig:e2e_results}
    \end{figure*}

    As a further validation, we apply the full correction stack to a set of synthetic observations. These observations have Gaussian random line amplitudes, Doppler shifts, and line widths, with some smoothing so that the overall intensity pattern resembles the small-scale chromospheric features observed in the cooler SPICE lines. To these observations, we apply a SPICE-like PSF with non-zero angle in its orientation with respect to both the $x-\lambda$ and $y-\lambda$ planes, so that it has non-zero shifts in both x and y. We synthesize SPICE-equivalent data with this PSF, then search for its parameters with our algorithm, producing line fits at the end. For a fully apples-to-apples comparison, we produce another SPICE-equivalent data set using a PSF with no $x-\lambda$ or $y-\lambda$ orientation, so that this data set has no PSF-induced Doppler artifacts. The results of this `end-to-end' test are shown in Figure~\ref{fig:e2e_results}. We find that the residual differences between the corrected and no PSF aberration Doppler shift datasets features are less than 5\,km/s, except in very low signal regions. This agreement is shown in the bottom left of Figure~\ref{fig:e2e_results}.

\section{Results \& Discussion}
\label{sec:Results}
    
     We have described a new method to derive Doppler velocities, and more generally to remove a persistent optical aberration, in the data from the SolO/SPICE instrument. The new method removes the PSF artifacts in the data, while applying a skew correction of the SPICE spectral cubes. We have tested the method on synthetic dataset in an end-to-end instrumental fashion to show that the method works as expected and the results are shown in Figure~\ref{fig:e2e_results}. Furthermore, comparing the results from our method on real SPICE data with corresponding IRIS observations, shown in Figures~\ref{fig:initial_psf_issue_illustration_Doppler}, illustrate that our method works as anticipated.
     
\begin{figure*}[!htbp]
        \begin{center}
            \includegraphics[width=0.75\textwidth]{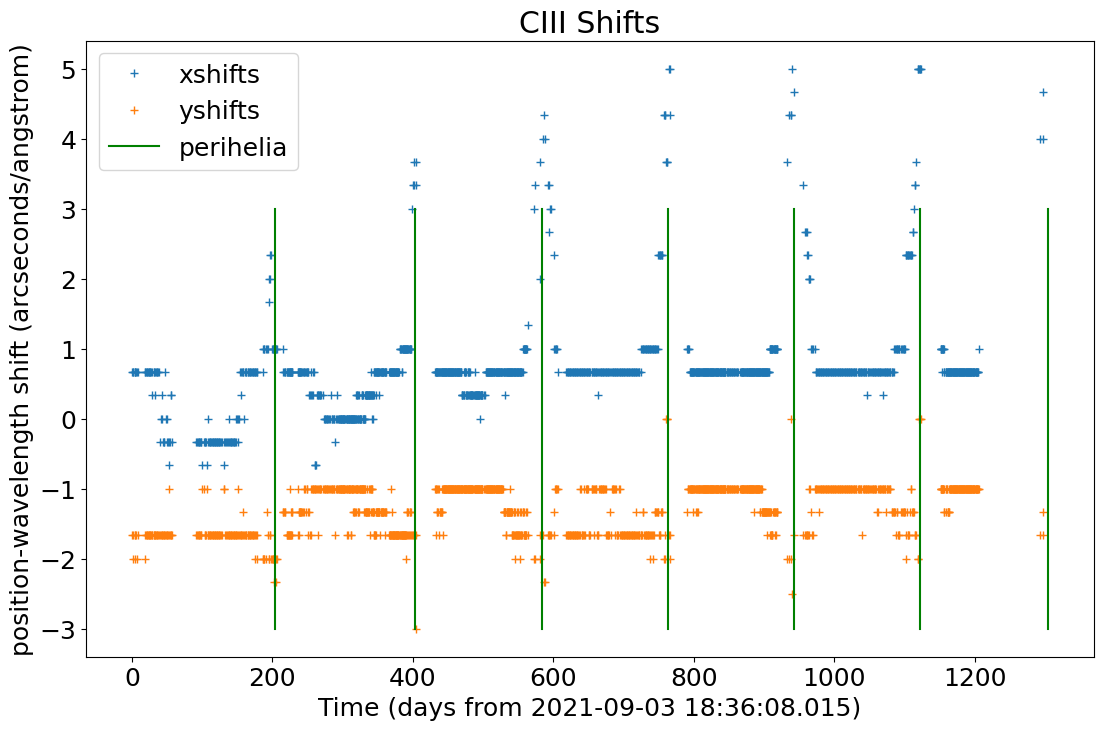}
            \includegraphics[width=0.75\textwidth]{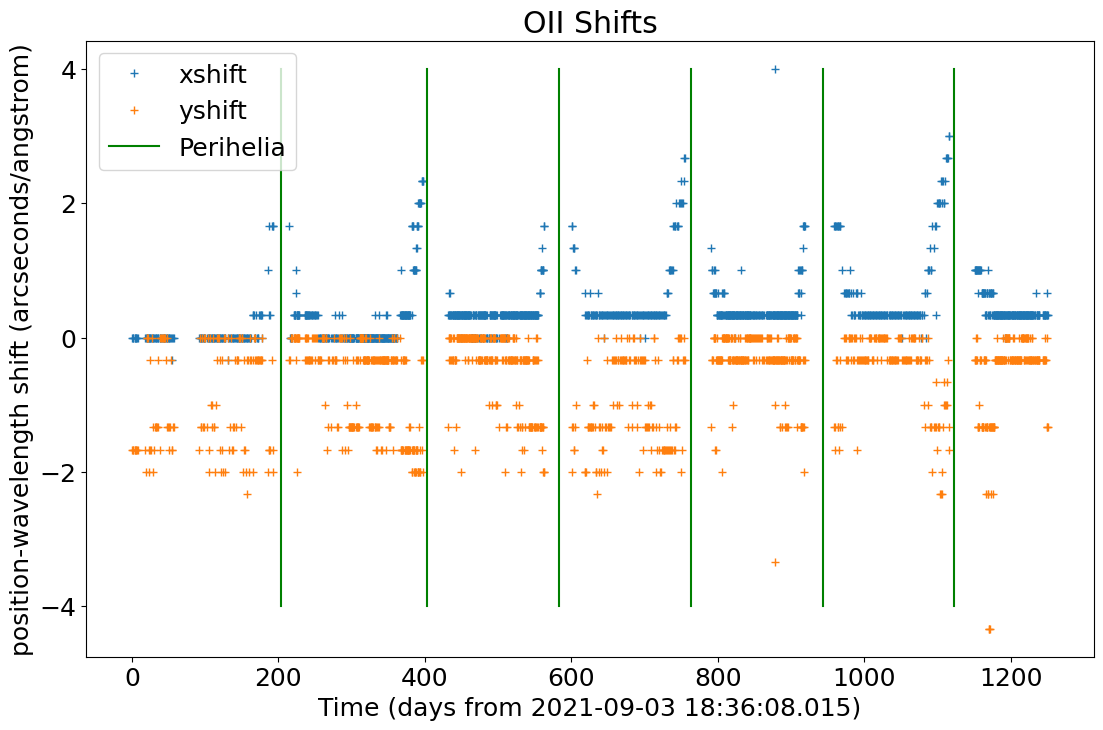}
        \end{center}
        \caption{Time variation of the $x$ and $y$ shift parameters computed in the SPICE SOUP observations. The upper panel shows the shifts for the C III 977\,\AA\ line; the lower panel shows the same for the O II 718\,\AA\, which is a relatively dim line in the shortwave detector. The shifts show large changes near perihelia passages, and return to a consistent background level (outside of the first $\sim 2$ years) when the spacecraft is more distant from the Sun. The scatter in the shifts is a rough indicator of the level of uncertainty in the observations. The shifts differ between the spectral lines, indicating that there is a wavelength or detector-position dependence in the amount of shift, although some of the difference could be due to lower SNR in the O II line. The green vertical lines show the SolO perihelia times. 
        }
        \label{fig:time_variability}
    \end{figure*}

     We must emphasize that this method can operate on data from time ranges when the SPICE PSF is not known \textit{a priori}. Hence, an investigation of the time evolution of the shift parameters is warranted. We have derived the shifts, as measured by our search algorithm for the SPROUTs data set, which is a SPICE synoptic observing program ~\citep{2025arXiv250212045V}. The SPROUTs data set contains a wide variety of useful spectral lines, with spectral rasters taken daily -- except during remote sensing windows where coordinated observing programs are run instead. We have included data from the same spectral lines from other observing programs during the important perihelia times, when the SPROUTs program was not being run. In particular, we concentrate on two spectral lines positioned on each of the detectors -- the C III 977 {\AA} line, top panel of Figure~\ref{fig:time_variability}, and the O II 718 \AA\ line in the bottom panel. The time variability of the shift parameters is shown in Figure~\ref{fig:time_variability}. The shifts vary dramatically near the perihelia (shown as the green vertical lines), but are largely static at other times, returning to similar equilibrium values at cruising distances. The results also suggest that there may be some differences between the two spectral lines compared, hinting a wavelength dependence of the phenomenon, although this could in part be because of lower signal-to-noise ratio (SNR) in the O II line.

\begin{figure*}[!htbp]
        \begin{center}
            \includegraphics[width=\textwidth]{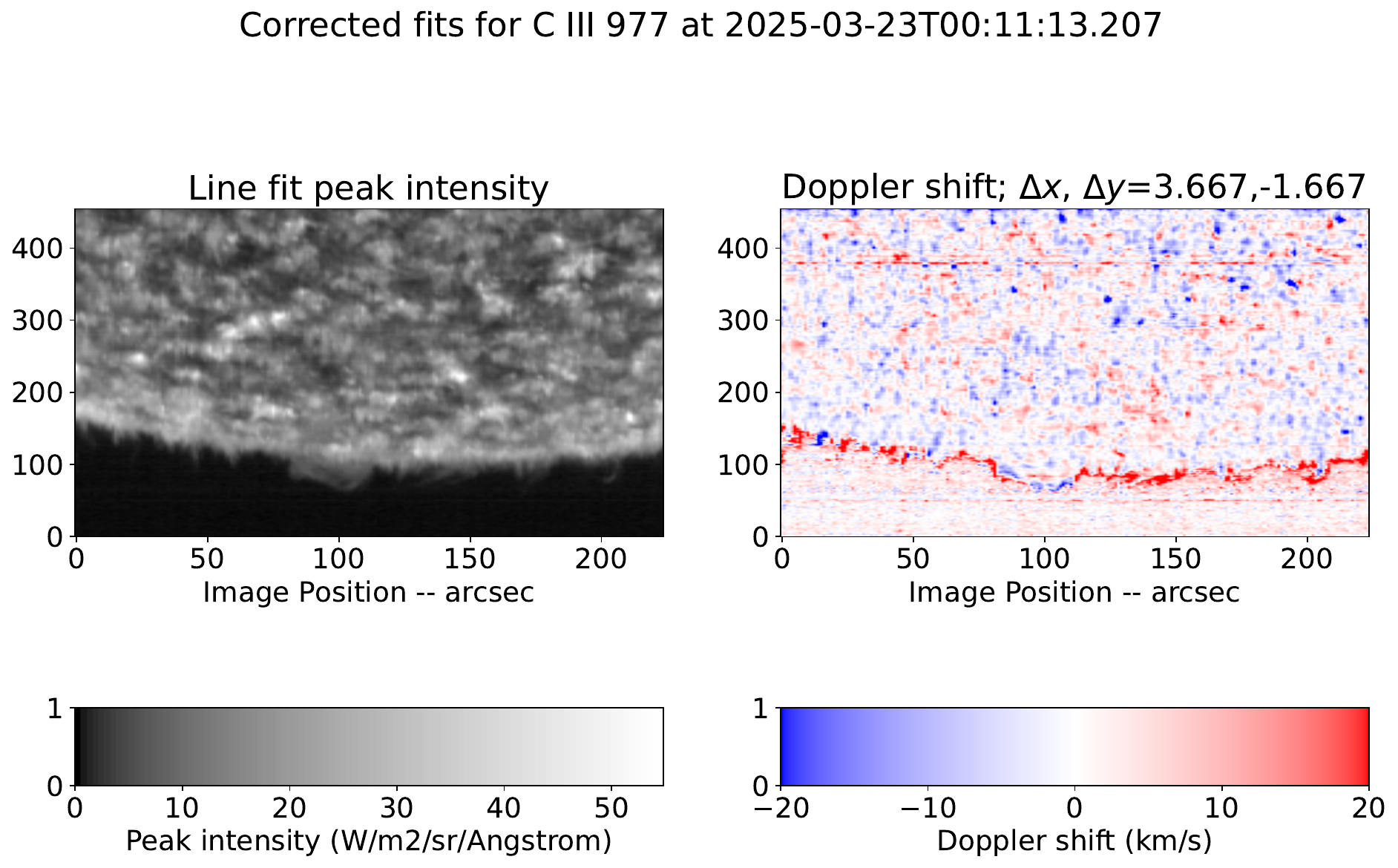}
        \end{center}
        \caption{ Example polar observations with PSF corrections for data taken on March 23, 2025. The corrections allow high quality Doppler shift measurements to be taken at the Sun's poles, never measured before, as Solar Orbiter progresses to higher latitude orbits in later phases of its mission.}\label{fig:polar_doppler}
    \end{figure*}

     To illustrate the capabilities of the new method, we apply it to the novel observations from Orbiter's recent high-latitude passage of the solar poles --  Doppler shift maps from our processing algorithm for the C III 977 \AA\  spectral line are shown in Figure~\ref{fig:polar_doppler}. These have never before been obtained, as no previous observatories with Doppler imaging have reached the high latitudes of Solar Orbiter. These observations were made on 23 March 2025, between 00:11 and 03:55 UT. The exposure time was $\sim$60 seconds, and a 4-arcsecond slit was used for raster scan with 224 steps. The Stonyhurst latitude of the Solar Orbiter was $\sim$ -17 degrees providing the high latitude perspective of the polar regions. Previous solar spectroscopic observations have had only the most oblique view of the poles, but now these SPICE observations will allow scientists to peer down onto the unique solar polar regions which will be the subject of future inquiries.  

    It is important to highlight that this correction requires raster data to correct the $x-\lambda$ shift. Sit-and-stare data does not contain the necessary information from outside the slit to correct the PSF effects, except purely those that are within the $y-\lambda$ plane. Likewise for `picket fence' mode of SPICE observations, where the stepping of the slit raster is larger than the size of the slit. The raster sampling should be comparable to the size of the PSF or less. Provided this constraint is satisfied, a narrow raster scan (5-10 positions) should be sufficient to allow a correction. Furthermore, the results of the temporal dependence of the shifts are a diagnostic of when the sit-and-stare data can be corrected with this or the previous algorithm -- specifically, when the $x$-shift is close to being zero. This inferrence is crucial for the proper analysis of Doppler velocity signals from SPICE sit-and-stare observations.
    
    We have made the code available on Github\footnote{\url{https://github.com/jeplowman/spice-line-fits}} \citep{spice_line_fit_doi}, along with a variety of other pieces of code. This codebase contains everything required to make the method work, but may also be very useful for other SPICE data analysis applications. In particular, the codebase includes:

    \begin{description}
        \item[\textbf{Doppler artifact correction:}] 
        The code to apply the skew correction to a set of SPICE data (typically Level 2 raster data), and to remove the residual spatial shift introduced by the correction, from spectral line fits -- i.e., the final `deskewing' step shown in Figure \ref{fig:Algorithm_figure}. 
        \item[\textbf{Correction parameter estimation:}] Code to estimate the best correction parameters using the methods described above.
        \item[\textbf{Line fitting:}] Code to fit spectral line profile(s) to a SPICE window. This code can fit single or multiple lines, using a Gaussian profile with a flat continuum level. The correction code is designed to be modular with respect to line fitting codes, however, so other methods (e.g., those using non-Gaussian line profiles) can be substituted.
        \item[\textbf{Spectral line fit uncertainty estimation:}] Code to estimate the uncertainties in the spectral line fits. The code is based on the Jacobian of the line fits at the best fitting parameters (see Appendix \ref{App:Jacobi_errors}). This is integral with the line fitting and so is technically part of the line fitting package, however, we list it separately since we feel it's an important innovation.
        \item[\textbf{Spatial Detrending:}] Code to remove a linear trend in the Doppler maps from spectral line fits results.
        \item[\textbf{Line fit result storage:}] This is code to place the results of the spectral line fits in standard .fits files, with relevant headers preserved, and header information that is human readable and relatable to the original SPICE data. This code interfaces with the other packages and it is intended to allow easier interoperability with other line fitting codes and other analysis software environments (e.g., Python vs. IDL).
    \end{description}

    Correcting SPICE's PSF is essential for enhancing the diagnostic capabilities of its observations. The method presented here offers a robust and improved correction of the $x-\lambda$ and $y-\lambda$ distortions, overcoming limitations of the previous matrix inversion technique and leading to more accurate velocity maps, without requiring prior knowledge of the PSF.
    These refined velocity maps will not only improve the interpretation of the elemental composition maps but also provide insights into plasma fractionation mechanisms \citep{Mondal_2021, to_spatially_2024}. Furthermore, accessing the direction and speed of plasma flows will help discriminate between competing theoretical models and trace back the source regions of the solar wind.



 
\begin{acknowledgements}
Solar Orbiter is a space mission of international collaboration between ESA and NASA, operated by ESA. The development of the SPICE instrument was funded by ESA and ESA member states (France, Germany, Norway, Switzerland, United Kingdom). The SPICE hardware consortium was led by Science and Technology Facilities Council (STFC) RAL Space and included Institut d’Astrophysique Spatiale (IAS), Max-Planck-Institut f\"ur Sonnensystemforschung (MPS), Physikalisch-Meteorologisches Observatorium Davos and World Radiation Center (PMOD/WRC), Institute of Theoretical Astrophysics (University of Oslo), NASA Goddard Space Flight Center (GSFC) and Southwest Research Institute (SwRI). The effort at SwRI for Solar Orbiter SPICE are supported by NASA under GSFC subcontract \#80GSFC20C0053 to Southwest Research Institute. 
\end{acknowledgements}

\bibliographystyle{aa}
\bibliography{psfrestore_paper1}

\newpage
    
\begin{appendix}
    \section{Jacobian error estimate for Least Squares Fitting of Spectral Lines}\label{App:Jacobi_errors}\vspace{-7cm}
    We define the Jacobian $J_{ij}$ to be the gradient of the residual vector $r_j$ with respect to the parameters of the solution $x_i$ (this is also the form returned by \texttt{scipy.optimize.least\_squares}):
    \begin{equation}
        J_{ij} \equiv \frac{\partial r_i}{\partial x_j}
    \end{equation}
    Where the residual vector is 
    \begin{equation}
        r_j = \frac{D_j - M_j(\mathbf{x_j)}}{\sigma_j}
    \end{equation}
    with data elements $D_j$ (in the case considered in this paper, they are the measured spectrum at a given $x-y$ position in the datacube, and $j$ is the wavelength pixel index), model (fitting) function $M_j$ (in this case a Gaussian with a background/continuum level), and the uncertainties in $D_j$ are $\sigma_j$. It follows that the $\chi^2$ merit function which least squares fitting attempts to minimize (WRT the model parameter vector $\mathbf{x}$) is simply 
    \begin{equation}
        \chi^2 = \sum_j r_j^2.
    \end{equation}
    When $\chi^2$ is equal to the number of degrees of freedom, ($n_f\equiv n_d-n_p$, with $n_d$ data points and $n_p$ parameters in the fit), the data is fully fit by the model and further improvements to the fit are not statistically significant. The quantity $\chi_r^2=\chi^2/n_f$ is called the `reduced' chi squared.\vspace{-6.75cm}

    We seek to estimate the uncertainties in the model fit parameters. That is, `How much can the fit parameters change, compared to the best fit, before the change results in a statistically significant difference in the merit function?' We define statistically significant to mean that the reduced $\chi^2$ between the best fit and a new set of parameters is one. We will further make the simplifying assumption that the parameters are uncorrelated in their effect on the residuals so that we can treat their errors independently. We also assume that the errors are small so that we can linearly expand about the best fit parameters, $\mathbf{x_0}$. In that case the parameter change vector, $\mathbf{\sigma_x}$, required to be statistically significant is easily expressed in terms of the Jacobian. $\chi_r^2=1$ gives the following starting equation:
    \begin{equation}\label{eq:uncertainty_criterion}
        \frac{1}{n_f}\sum_j\frac{[M_j(\mathbf{x_0})-M_j(\mathbf{x_0}-\mathbf{\sigma_x})]^2}{\sigma_j^2}=1
    \end{equation}
    Taylor expanding about $\mathbf{x_0}$, for model parameter $i$ (with uncertainty $\sigma_{xi}$) we have 
    \begin{equation}
        \frac{M_j(\mathbf{x_0}) - M_j(\mathbf(x_0)-\mathbf{\sigma_x})}{\sigma_j} = \sigma_{xi}\frac{\partial M_j}{\partial x_i}\bigg\rvert_{\mathbf{x_0}} = \sigma_{xi}J_{ij}\big\rvert_{\mathbf{x_0}}
    \end{equation}
    Equation \ref{eq:uncertainty_criterion} therefore yields
    \begin{equation}
        \sigma_{xi}^2 = \frac{n_f}{\sum_j J_{ij}^2}.
    \end{equation}
    With the assumption of uncorrelated error contributions from each parameter, their individual contributions to the figure of merit add in quadrature. The final estimate per parameter, with all parameters considered, is therefore
    \begin{equation}
        \sigma_{xi} = \sqrt{\frac{n_f}{n_p\sum_{ij} J_{ij}^2}}.
    \end{equation}
    This is the basis of the uncertainty estimate included by our line fitting code.
\end{appendix}

\end{document}